\documentclass{article}

\usepackage{microtype}
\usepackage{graphicx}
\usepackage{subcaption}
\usepackage{booktabs} 
\usepackage{amsmath}
\usepackage{amssymb}
\usepackage{adjustbox}
\usepackage{xcolor}
\usepackage{multirow}
\usepackage{booktabs}
\usepackage{romannum}
\usepackage{multicol}

\newcommand\norm[1]{\left\lVert#1\right\rVert}
\newcommand{\E}{\mathbb{E}}

\usepackage{hyperref}

\usepackage[accepted]{icml2021}

\icmltitlerunning{Meta-StyleSpeech : Multi-Speaker Adaptive Text-to-Speech Generation}

\begin{document}

\twocolumn[
\icmltitle{Meta-StyleSpeech : Multi-Speaker Adaptive Text-to-Speech Generation}




\begin{icmlauthorlist}
\icmlauthor{Dongchan Min}{KA}
\icmlauthor{Dong Bok Lee}{KA}
\icmlauthor{Eunho Yang}{KA,AI}
\icmlauthor{Sung Ju Hwang}{KA,AI}
\end{icmlauthorlist}

\icmlaffiliation{KA}{Graduate School of AI, Korea Advanced Institute of Science and Technology (KAIST), Seoul, South Korea}
\icmlaffiliation{AI}{AITRICS, Seoul, South Korea}

\icmlcorrespondingauthor{Dongchan Min}{alsehdcks95@kaist.ac.kr}
\icmlcorrespondingauthor{Sung Ju Hwang}{sjhwang82@kaist.ac.kr}

\icmlkeywords{Machine Learning, ICML, TTS, Text-to-Speech, Speech, MultiSpeeker TTS}

\vskip 0.3in
]



\printAffiliationsAndNotice{}  

\begin{abstract}
With rapid progress in neural text-to-speech (TTS) models, personalized speech generation is now in high demand for many applications. For practical applicability, a TTS model should generate high-quality speech with only a few audio samples from the given speaker, that are also short in length. However, existing methods either require to fine-tune the model or achieve low adaptation quality without fine-tuning. In this work, we propose StyleSpeech, a new TTS model which not only synthesizes high-quality speech but also effectively adapts to new speakers. Specifically, we propose Style-Adaptive Layer Normalization (SALN) which aligns gain and bias of the text input according to the style extracted from a reference speech audio. With SALN, our model effectively synthesizes speech in the style of the target speaker even from a single speech audio. Furthermore, to enhance StyleSpeech's adaptation to speech from new speakers, we extend it to Meta-StyleSpeech by introducing two discriminators trained with style prototypes, and performing episodic training. The experimental results show that our models generate high-quality speech which accurately follows the speaker's voice with single short-duration (1-3 sec) speech audio, significantly outperforming baselines.
\end{abstract}

\section{Introduction}

In the past few years, the fidelity and intelligibility of speech produced by neural text-to-speech (TTS) synthesis models have shown dramatic improvements. Furthermore, a number of applications such as AI voice assistant services and audio navigation systems have been actively developed and deployed to real-world, attracting increasing demand of TTS synthesis. The majority of the TTS models aim to synthesize high quality speech of a single speaker from the given text~\citep{Oord2016WaveNetAG, Wang2017TacotronTE, Shen2018NaturalTS, Ren2019FastSpeechFR, Ren2020FastSpeech2F} and have been extended to support multi speakers \citep{Gibiansky2017DeepV2, Ping2017DeepV3, Chen2020MultiSpeechMT}.

Meanwhile, there is an increasing demand for \emph{personalized speech generation}, which requires TTS models to generate high-quality speech that well captures the voice of the given speaker with only a few samples of the speech data. However, natural human speech is highly expressive and contains rich information, including various factors such as the speaker identity and prosody. Thus, generating personalized speech from a few speech audios, potentially even from a single audio, is an extremely challenging task. To achieve this goal, the TTS model should be able to generate speech of multiple speakers and adapt well to an unseen speaker's voice.

A popular approach to handle this challenge is to pre-train the model on a large dataset consisting of the speech from many speakers and fine-tune the model with a few audio samples of a target speaker \citep{Chen2019SampleEA, Arik2018NeuralVC, chen2021adaspeech}. However, this approach requires audio samples and corresponding transcripts of the target speaker to fine-tune, as well as hundreds of fine-tuning steps, which limits its applicability to real-world scenarios. Another approach is to use a piece of reference speech audio to extract a latent vector that captures the voice of the speaker such as speaker identity, prosody and speaking style \cite{SkerryRyan2018TowardsEP, Wang2018StyleTU, Jia2018TransferLF}. Specifically, these models are trained to synthesize speech conditioned on a \emph{latent style vector} extracted from the speech of the given speaker, in addition to text input. This style-based approach has shown convincing results in expressive speech generation, and is able to adapt to new speakers without fine-tuning. However, they heavily rely on the diversity of the speakers in the source dataset, and thus often show low adaptation performance on new speakers. 

Meta-learning \citep{learningtolearn}, or learning to learn, has attracted attention of many researchers recently, as it allows the trained model to rapidly adapt to new tasks only with a few examples. In this regard, some of the few-shot generative models have utilized meta-learning for improved generalization \citep{Rezende2016OneShotGI, Bartunov2018FewshotGM, Cloutre2019FIGRFI}. Closely related to few-shot generation, few-shot classification also is the most extensively studied problem for meta-learning. In particular, metric-based methods \citep{Snell2017PrototypicalNF, Misra2020MishAS}, which meta-learn a space where the instances from the same class are embedded closer while instances that belong to different classes are embedded farther apart, have shown to achieve high performance. However, the existing methods on few-shot generation and classification mostly targets image domains, and are not straightforwardly applicable to TTS.

To overcome these difficulties, we propose \emph{StyleSpeech}, a high-quality and expressive multi-speaker adaptive text-to-speech generation model. Our model is inspired by \citet{Karras2019ASG} proposed for image generation, that are shown to generate surprisingly realistic photos of human faces. Specifically, we propose \emph{Style-Adaptive Layer Normalization} (SALN) which aligns gain and bias of the text input according to a style vector extracted from a reference speech audio. Additionally, we further propose \emph{Meta-StyleSpeech}, which is \emph{meta-learned} with discriminators to further improve the model's ability to adapt to new speakers that have not been seen during training. In particular, we perform episodic training by simulating the one-shot adaptation case in each episode, while additionally training two discriminators, a style discriminator and a phoneme discriminator with adversarial loss. Furthermore, the style discriminator learns a set of style prototypes enforcing the generator to generate speech from each speaker to be embedded closer to its correct style prototype which can be a voice identity of the speaker. 


Our main contributions are as follows:
\begin{itemize}
\item We propose \emph{StyleSpeech}, a high-quality and expressive multi-speaker adaptive TTS model, which can flexibly synthesize speech with the style extracted from a single short-length reference speech audio.
\item We extend StyleSpeech to \emph{Meta-StyleSpeech} which adapts well to speech from unseen speakers, by introducing the phoneme and style discriminators with style prototypes and an episodic meta-learning algorithm 
\item Our proposed models achieve start-of-the-art TTS performance across multiple tasks, including multi-speaker speech generation and one-shot short-length speaker adaptation.
\end{itemize}


\section{Related Work}

\paragraph{Text-to-Speech} 
Neural TTS models have shown a rapid progress, including WaveNet \citep{Oord2016WaveNetAG}, DeepVoice1, 2, 3 \citep{Arik2017DeepVR, Gibiansky2017DeepV2, Ping2017DeepV3}, Char2Wav \citep{Sotelo2017Char2WavES} and Tacotron1, 2 \citep{Wang2017TacotronTE, Shen2018NaturalTS}. These models mostly resort to autoregressive generation of mel-spectrogram, which suffers from slow inference speed and a lack of robustness (word missing and skipping). Recently, several works such as Paranet \citep{Peng2020NonAutoregressiveNT} and FastSpeech1 \citep{Ren2019FastSpeechFR}
have proposed non-autoregressive TTS models to handle such issues and achieve fast inference speed and improved robustness over autoregressive models. Besides, FastSpeech2 \citep{Ren2020FastSpeech2F} extend FastSpeech1 by using additional pre-obtained acoustic features such as pitch and energy so that they show more expressive speech generation. However, these models only support a single speaker system. In this work, we base our model on FastSpeech2 and we propose an additional component to generate various voice of multi speakers.

\paragraph{Speaker Adaptation} 
As the demand for personalized speech generation have increased, adaptation of the TTS models to new speakers has been extensively studied. A popular approach is to train the model on a large multi-speakers dataset and then fine-tune the whole model \citep{Chen2019SampleEA, Arik2018NeuralVC} or only parts of the model \citep{Moss2020BOFFINTF, Zhang2020AdaDurIANFA, chen2021adaspeech}. As an alternative approach, some of recent works attempted to model the style directly from a speech audio sample. For example, \citet{Nachmani2018FittingNS} extend VoiceLoop \citep{Taigman2018VoiceLoopVF} with an additional speaker encoder. Tacotron-based approaches \citep{SkerryRyan2018TowardsEP, Wang2018StyleTU} use a piece of reference speech audio to extract the style vector and synthesize speech with the style vector, in addition to text input. Moreover, GMVAE-Tacotron \citep{Hsu2019HierarchicalGM} present a variational approach with Gaussian Mixture prior in style modeling. To certain extent, these methods appear to be able to generate speech, that are not limited to trained speakers. However, they often achieve low adaptation performance on unseen speakers, especially when the reference speech is short in length. To handle this issue, we introduce a novel method to flexibly synthesize speech with the style vector and propose meta-learning to improve adaptation performance of our model on unseen speakers.


\paragraph{Meta Learning} 
In recent years, diverse meta-learning algorithms have been proposed, mostly focusing on the few-shot classification problem. Among them, metric-based meta-learning \citep{metric2, Snell2017PrototypicalNF, metric4} aims to learn the embedding space where the instances of the same class are embedded closer, while the instances belonging to different classes are embedded farther apart. On the other hand, some of existing works have studied the problem of few-shot generation. For one-shot image generalization, \citet{Rezende2016OneShotGI} propose sequential generative models that are built on the principles of feedback and the attention mechanism. \citet{Bartunov2018FewshotGM} developed a hierarchical Variational Autoencoder for few-shot image generation, while \citet{Reed2018FewshotAD} propose the extension of Pixel-CNN with neural attention for few-shot auto-regressive density modeling. However, all these methods focus on the image domain. In contrast, our work focuses on the few-shot, short-length adaptation of TTS models, which has been relatively overlooked.

\begin{figure}[t]
\begin{center}
\centering
{\includegraphics[width=0.95\columnwidth]{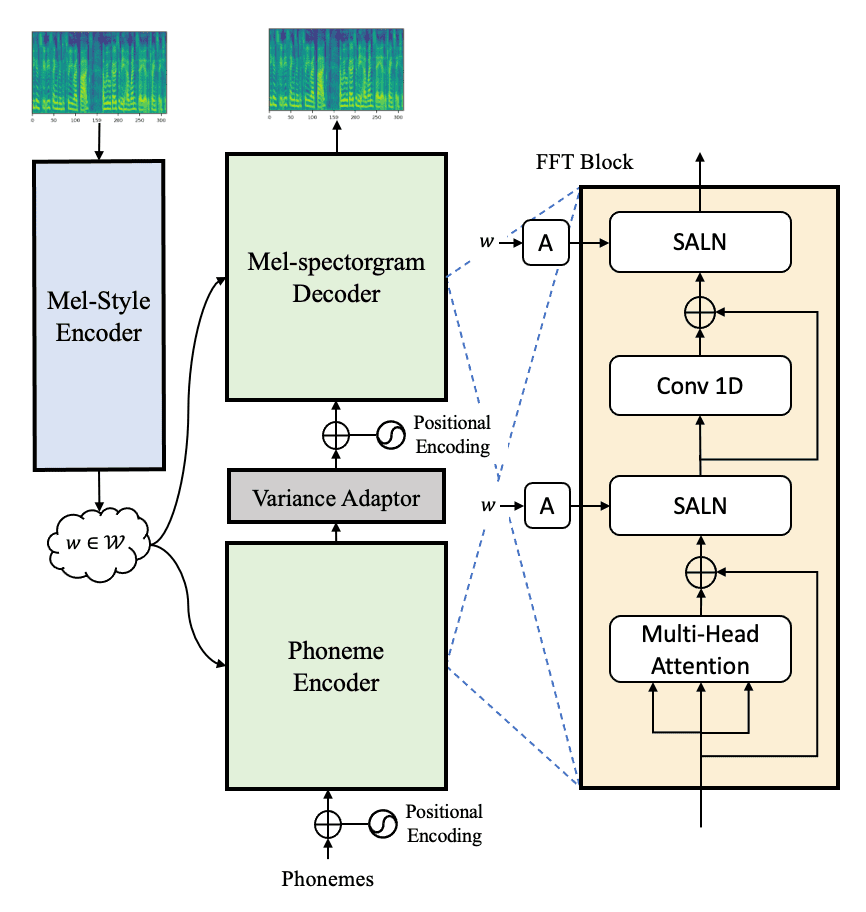}}
\vskip -0.1in
\caption{\small \textbf{The architecture of StyleSpseech.} The mel-style encoder extracts the style vector from a reference speech sample, and the generator converts the phoneme sequence into speech of various voices through the Style-Adaptive Layer Normalization (SALN).}
\label{figure1}
\end{center}
\vskip -0.3in
\end{figure}

\begin{figure*}[t]
\centering
\includegraphics[width=0.88\textwidth]{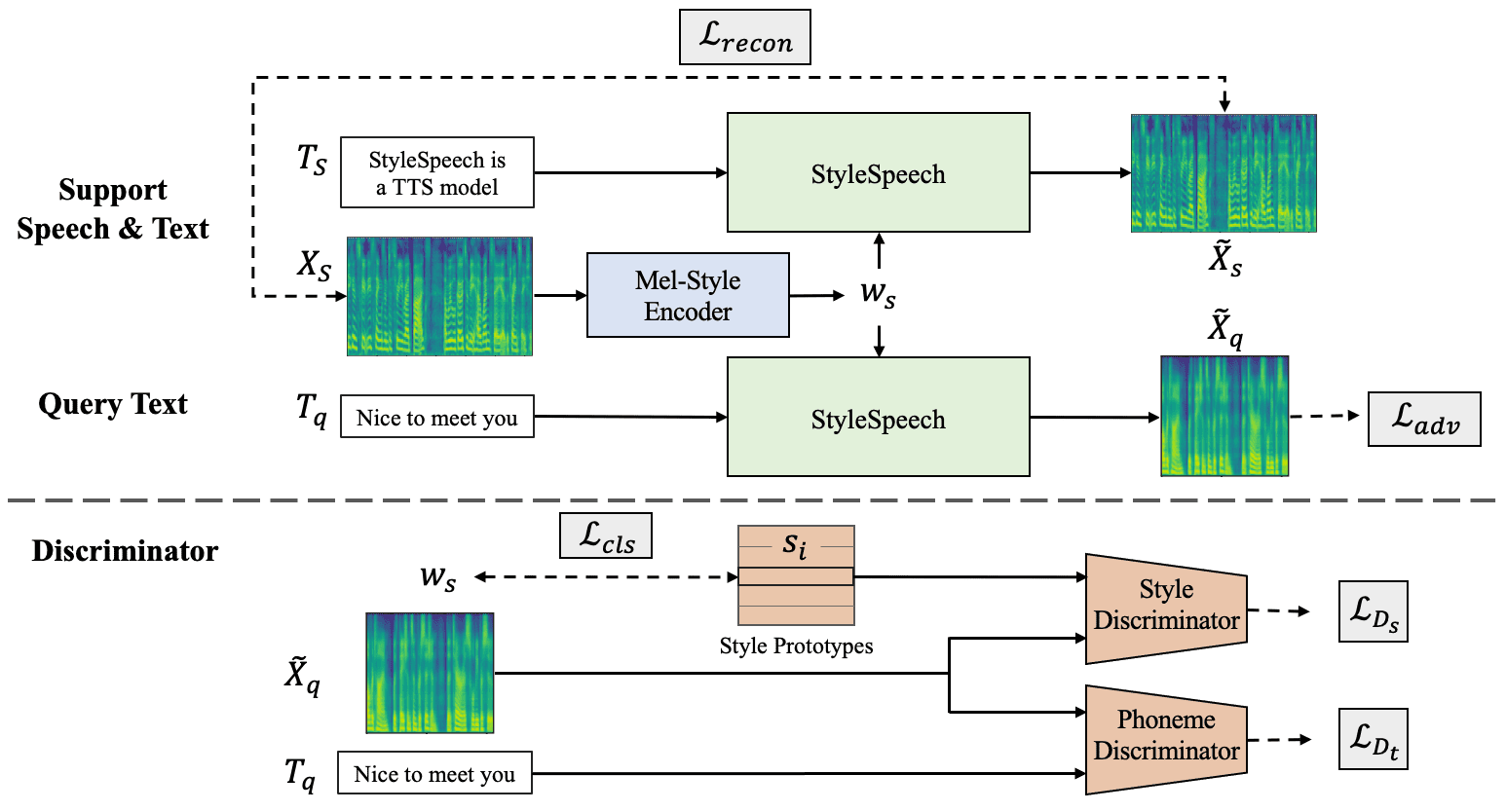}
\caption{\small \textbf{Overview of Meta-StyleSpeech.} The generator use the style vector extracted from support speech and the query text to synthesize the query speech. The style discriminator learns a set of style prototypes for each speaker and enforces the generated speech to be gathered around the style prototype of the target speaker. The phoneme discriminator distinguish the real speech and generated speech condition on the input text.}
\label{figure2}
\vskip -0.1in
\end{figure*}


\section{StyleSpeech}
In this section, we first describe the architecture of StyleSpeech for multi-speaker speech generation. StyleSpeech is comprised of a mel-style encoder and a generator. The overall architecture of StyleSpeech is shown in Figure \ref{figure1}.

\subsection{Mel Style Encoder}
The \emph{mel-style encoder}, $Enc_s$, takes a reference speech $X$ as input. The goal of the mel-style encoder is to extract a vector $w \in \mathbb{R}^N $ which contains the style such as speaker identity and prosody of given speech $X$. Similar to \citet{Arik2018NeuralVC}, we design the mel-style encoder to comprise of the following three parts: 

1) Spectral processing: We first input the mel-spectrogram into fully-connected layers to transform each frames of mel-spectrogram into hidden sequences.

2) Temporal processing: We then use gated CNNs \cite{Dauphin2017LanguageMW} with residual connection to capture the sequential information from the given speech. 

3) Multi-head self-attention: Then we apply a multi-head self-attention with residual connection to encode the global information. In contrast to \citet{Arik2018NeuralVC} where the multi-head self attention is applied across audio samples, we apply it at the frame level so that the mel-style encoder can better extract style information even from a short speech sample. Then, we temporally average the output of the self-attention to get a one-dimensional style vector $w$.

\subsection{Generator}
The \emph{generator}, $G$, aims to generate a speech $\widetilde{X}$ given a phoneme (or text) sequence $t$ and a style vector $w$. We build the base generator architecture upon FastSpeech2~\citep{Ren2020FastSpeech2F}, which is one of the most popular single-speaker models in non-autoregressive TTS. The model consists of three parts; a phoneme encoder, a mel-spectrogram decoder and a variance adaptor. The phoneme encoder converts a sequence of phoneme embedding into a hidden phoneme sequence. Then, the variance adaptor predicts different variances in the speech such as pitch and energy in phoneme-level \footnote{In multi-style generation, we find that it is straightforward to predict the phoneme-level information than speech frame-level information as in FastSpeech2.}. Furthermore, the variance adaptor predict a duration of each phonemes to regulate the length of the hidden phoneme sequence into the length of speech frames. Finally, the mel-spectrogram decoder converts the length-regulated phoneme hidden sequence into mel-spectrogram sequence. Both the phoneme encoder and mel-spectrogram decoder are composed of Feed-Forward Transformer blocks (FFT blocks) based on the Transformer \citep{Vaswani2017AttentionIA} architecture. However, this model does not generate speech with diverse speakers, and thus we propose a novel component to support multi-speaker speech generation, in the following paragraph.

\textbf{Style-Adaptive Layer Norm} \; Conventionally, the style vector is provided to the generator simply through either the concatenation or the summation with the encoder output or the decoder input. In contrast, we apply an alternative approach by proposing the \textit{Style-Adaptive Layer Normalization} (SALN). SALN receives the style vector $w$ and predicts the gain and bias of the input feature vector. More precisely, given feature vector $\boldsymbol{h} = (h_1, h_2, \dots, h_H)$ where $H$ is the dimensionality of the vector, we derive the normalized vector $\boldsymbol{y}=(y_1, y_2, \dots, y_H)$ as follows:
\begin{equation}
\begin{gathered}
\boldsymbol{y} = \frac{\boldsymbol{h} - \mu}{\sigma} \\ 
\mu = \frac{1}{H} \sum^H_{i=1}\, h_i, \quad \sigma = \sqrt{\frac{1}{H}\sum^H_{i=1}(h_i-\mu)^2}
\end{gathered}
\end{equation}
Then, we compute the gain and bias with respect to the style vector $w$.
\begin{equation}
SALN(\boldsymbol{h}, w) = g(w) \cdot \boldsymbol{y} + b(w)
\end{equation}
Unlike the fixed gain and bias as in LayerNorm \citep{Ba2016LayerN}, $g(w)$ and $b(w)$ can adaptively perform scaling and shifting of the normalized input features based on the style vector. We substitute SALN for layer normalizations in FFT blocks in the phoneme encoder and the mel-spectrogram decoder. The affine layer which convert the style vector into bias and gain is a single fully connected layer. By utilizing SALN, the generator can synthesize various styles of speech of multiple speakers given the reference audio sample in addition to the phoneme input.

\subsection{Training}
In the training process, both the generator and the mel-style encoder are optimized by minimizing a reconstruction loss between a mel-spectrogram synthesized by the generator and a ground truth mel-spectrogram\footnote{The reconstruction loss includes pitch, energy, and duration losses as in FastSpeech2, however, we do not present them for brevity.}. We use the $L_1$ distance as a loss function, as follows:

\begin{equation}
   \widetilde{X} = G(t, w) \quad  w = Enc_s(X)
\end{equation}
\begin{equation}
    \mathcal{L}_{recon} = \E\left[ \left\| \widetilde{X} - X\right\|_1 \right]
\end{equation}

where $\widetilde{X}$ is a generated mel-spectrogram given the phoneme input, $t$, and the style vector, $w$, which extracted from a ground truth mel-spectrogram, $X$.


\section{Meta-StyleSpeech}
Although StyleSpeech can adapt to the speech from a new speaker by utilizing SALN, it may not generalize well to the speech from an unseen speaker with a shifted distribution. Furthermore, it is difficult to generate the speech to follow the voice of the unseen speaker, especially with few speech audio samples that are also short in length. Thus, we further propose \emph{Meta-StyleSpeech}, which is meta-learned to further improve the model's ability to adapt to unseen speakers. In particular, we assume that only a single speech audio sample of the target speaker is available. Thus, we simulate one-shot learning for new speakers via episodic training. In each episode, we randomly sample one support (speech, text) sample, $(X_s, t_s)$, and one query text, $t_q$, from the target speaker $i$. Our goal is then to generate the query speech $\widetilde{X}_q$ from the query text $t_q$ and the style vector $w_s$ which is extracted from the support speech $X_s$. However, a challenge here is that we can not apply reconstruction loss on $\widetilde{X}_q$, since no ground-truth mel-spectrogram is available. To handle this issue, we introduce an additional adversarial network with two discriminators; a style discriminator and a phoneme discriminator.

\subsection{Discriminators} 
The \emph{style discriminator}, $D_s$, predicts whether the speech follows the voice of the target speaker. The discriminator has similar architecture with mel-style encoder except it contains a set of \emph{style prototypes} $S=\{s_i\}_{i=1}^K$, where $s_i \in \mathbb{R}^N$ denotes the style prototype for the $i \,$th speaker and $K$ is the number of speakers in the training set. Given the style vector, $w_s \in \mathbb{R}^N$, as input, the style prototype $s_i$ is learned with following classification loss. 
\begin{equation}
    \mathcal{L}_{cls} = -\log \, \frac{\exp(w_s^T s_i))}{\sum_{i^{\prime}} \exp(w_s^T s_{i^{\prime}})}
\end{equation}
In detail, the dot product between the style vector and all style prototypes is computed to produce style logits, followed by cross entropy loss that encourages the style prototype to represent the target speaker's common style such as speaker identity.

The style discriminator then maps the generated speech $\widetilde{X}_q$ to a $M$-dimensional vector $h(\widetilde{X}_q) \in \mathbb{R}^M $ and compute a single scalar with the style prototype. The key idea here is to enforce the generated speech to be gathered around the style prototype for each speaker. In other words, the generator learns how to synthesize speech that follows the common style of the target speaker from a single short reference speech sample. Similar to the idea of \citet{Miyato2018cGANsWP}, the output of the style discriminator is then computed as:
\begin{equation}
    D_s(\widetilde{X}_q, s_i) = w_0 \, {s_{i}}^T V h(\widetilde{X}_q) + b_0
\end{equation}
where $V \in \mathbb{R}^{N \times M}$ is a linear layer and $w_0$ and $b_0$ are learnable parameters. The discriminator loss function of $D_s$ then becomes
\begin{multline}
\small
    \mathcal{L}_{D_s} = \E_{t,w,s_i\sim S}[(D_s(X_s, s_i)-1)^2 + D_s(\widetilde{X}_q, s_i)^2]
\end{multline}
The discriminator loss follows LS-GAN \citep{Mao2017LeastSG}, which replace the binary cross-entropy terms of the original GAN \citep{Goodfellow2014GenerativeAN} objective with least squares loss functions. 

The \emph{phoneme discriminator}, $D_t$, takes $\widetilde{X}_q$ and $t_q$ as inputs, and distinguishes the real speech from the generated speech given the phoneme sequence $t_q$ as the condition. In particular, the phoneme discriminator consists of fully-connected layers and is applied at frame level. Since we know the duration of each phoneme, we can concatenate each frame in mel-spectrogram with the corresponding phoneme. The discriminator then computes scalars for each frame and averages them to get a single scalar. The final discriminator loss function for $D_t$ then is given as: 
\begin{multline}
\small
    \mathcal{L}_{D_t} = \E_{t,w}[(D_t(X_s, t_s)-1)^2 + D_t(\widetilde{X}_q, t_q)^2]
\end{multline}
Finally, the generator loss for query speech can be defined as the sum of the adversarial loss for each discriminator, as follows:
\begin{multline}
    \mathcal{L}_{adv} =
    \E_{t,w,s_i\sim S}[(D_s(G(t_q,w_s), s_i)-1)^2] + \\
    \E_{t,w}[(D_t(G(t_q,w_s), t_q)-1)^2] .
\end{multline}
Furthermore, we additionally apply a reconstruction loss for the support speech, as we empirically find that it improves the quality of the generated mel-spectrograms.
\begin{equation}
    \mathcal{L}_{recon} = \E\left[ \left\| G(t_s, w_s) - X_s\right\|_1 \right]
\end{equation}

\subsection{Episodic meta-learning}
Overall, we conduct the meta-learning of Meta-StyleSpeech by alternating between the updates of the generator and mel-style encoder that minimizes $\mathcal{L}_{recon}$ and $\mathcal{L}_{adv}$ losses, and the updates of the discriminators that minimizes $\mathcal{L}_{D_s}$, $\mathcal{L}_{D_t}$ and $\mathcal{L}_{cls}$ losses. Thus, the final meta-training loss to minimize is defined as :
\begin{equation}\label{Eq11}
    \mathcal{L}_{G} = \alpha \mathcal{L}_{recon} + \mathcal{L}_{adv}
\end{equation}
\begin{equation}\label{Eq12}
    \mathcal{L}_{D} =  \mathcal{L}_{D_s} + \mathcal{L}_{D_t} + \mathcal{L}_{cls}
\end{equation}
We set $\alpha=10$ in our experiments.

\section{Experiment}
In this section, we evaluate the effectiveness of Meta-StyleSpeech and StyleSpeech on few-shot text-to-speech synthesis tasks. The audio samples are available at \\ \url{https://stylespeech.github.io/}.

\subsection{Experimental Setup}

\paragraph{Datasets} We train StyleSpeech and Meta-StyleSpeech on \textbf{LibriTTS} dataset \citep{Zen2019LibriTTSAC}, which is a multi-speaker English corpus derived from LibriSpeech \citep{Panayotov2015LibrispeechAA}. LibriTTS contains 110 hours audios of 1141 speakers and their corresponding text transcripts. We split the dataset into a training and a validation (test) set, and use the validation set for the evaluation on the trained speakers. For evaluation of the models' performance on unseen speaker adaptation tasks, we use the \textbf{VCTK} \citep{Yamagishi2019CSTRVC} dataset which contains audios of 108 speakers.

\paragraph{Preprocessing} We convert the text sequences into the phoneme sequences with an open-source grapheme-to-phoneme tool\footnote{\url{https://github.com/Kyubyong/g2p}} and take the phoneme sequences as input. We downsampled an audio to 16kHz and trimmed leading and trailing silence using Librosa \citep{McFee2015librosaAA}. We extract a spectrogram with a FFT size of 1024, hop size of 256, and window size of 1024 samples. Then, we convert it to a mel-spectrogram with 80 frequency bins. In addition, we average ground-truth pitch and energy by duration to get phoneme-level pitch and energy.

\begin{table}[t]
\centering
\small
\begin{adjustbox}{width=0.97\linewidth}
\begin{tabular}{lccc}
\toprule
{\bf Model}	& {\bf MOS ($\mathbf{\uparrow}$)}	&  {\bf MCD ($\mathbf{\downarrow}$)} & {\bf WER ($\mathbf{\downarrow}$)} \\
\midrule

GT & 4.37$\pm$0.15 & - & - \\

GT \textit{mel} + Vocoder & 4.02$\pm$0.15 & - & - \\

\midrule
DeepVoice3 & 2.23$\pm$0.15 & 4.92$\pm$0.22 & 36.56  \\

GMVAE & 3.28$\pm$0.20 & 4.81$\pm$0.24 & 24.49  \\

{Multi-speaker FS2(\textit{vanilla})} & {3.53$\pm$0.13} & {4.50$\pm$0.21} &  {16.49} \\

{Multi-speaker FS2+\textit{d}-vector}  & {3.46$\pm$0.13} & {4.53$\pm$0.21} &  {16.51} \\

\midrule

StyleSpeech & 3.84$\pm$0.14 & 4.49$\pm$0.22 &  16.79 \\

\textbf{Meta-StyleSpeech} & \bf 3.89$\pm$0.12  & \bf 4.29$\pm$ \bf 0.21 & \bf 15.68 \\

\bottomrule
\end{tabular}
\end{adjustbox}
\caption{MOS, MCD and WER for seen speakers.}
\label{seen_mos}
\end{table}
\begin{table}[t]
\small
\centering
\begin{adjustbox}{width=0.73\linewidth}
\begin{tabular}{l|cc}
\toprule
\textbf{Model} & \textbf{SMOS ($\mathbf{\uparrow}$)} & \textbf{Sim ($\mathbf{\uparrow}$)}\\
\midrule
GT  & 4.75$\pm$0.14  & - \\
GT \textit{mel} + Vocoder & 4.57$\pm$0.14 & - \\
\midrule
DeepVoice3 & -  & 0.65  \\
GMVAE & 3.17$\pm$0.06  & 0.76  \\
Multi-speaker FS2(\textit{vanilla}) & 3.41$\pm$0.16  & 0.80  \\
Multi-speaker FS2+\textit{d}-vector & 2.42$\pm$0.08  & 0.66  \\
\midrule
StyleSpeech & 3.83$\pm$0.16 & 0.80  \\
\bf Meta-StyleSpeech & \bf 4.08$\pm$0.15 & \bf 0.81 \\
\bottomrule
\end{tabular}
\end{adjustbox}
\caption{SMOS and Sim for seen speakers.}
\label{seen_smos}
\vspace{-0.1in}
\end{table}

\paragraph{Implementation Details} The generator in StyleSpeech uses 4 FFT blocks on both phoneme encoder and mel-spectrogram decoder following FastSpeech2. In addition, we add pre-nets to the phoneme encoder and the mel-spectrogram decoder. In detail, the encoder pre-net consists of two convolution layers and a linear layer with residual connection and the decoder pre-net consists of two linear layers. The architecture of pitch, energy and duration predictor in the variance adaptor are the same as those of FastSpeech2. However, instead of using 256 bins for pitch and energy, we use a 1D convolution layer to add real or predicted pitch and energy directly to the phoneme encoder output \cite{Lancucki2020FastPitchPT}. For the mel-style encoder, the dimensionality of all latent hidden vectors are set to 128, including the size of the style vector. Furthermore, we use Mish \citep{Misra2020MishAS} activation for both the generator and the mel-style encoder. 

The style discriminator has the similar architecture as the mel-style encoder, but we use 1D convolutions instead of gated CNNs. The phoneme discriminator consist of fully-connected layers. Specifically, the mel-spectrogram passes through two fully connected layers with 256 hidden dimensions before concatenating with phoneme embeddings, and then through three fully connected layers with 512 hidden dimensions and one final projection layer. We use Leaky-ReLU activation for both discriminators. In addition, We apply spectral normalization \citep{Miyato2018SpectralNF} in all the layers for both discriminators except style prototypes. The more details about the architectures are in the supplementary material.

\begin{table}[t]
\centering
\begin{adjustbox}{width=0.97\linewidth}
\small
\begin{tabular}{lccc}
\toprule
{\bf Model}	& {\bf MOS ($\mathbf{\uparrow}$)}	&  {\bf MCD ($\mathbf{\downarrow}$)} & {\bf WER ($\mathbf{\downarrow}$)} \\
\midrule
GT & 4.40$\pm$0.13 & - & - \\
GT \textit{mel} + Vocoder & 4.03$\pm$0.12 & - & - \\
\midrule
GMVAE & 3.15$\pm$0.17 & 5.54$\pm$0.26 & 23.86  \\

{Multi-speaker FS2(\textit{vanilla})} & {3.69$\pm$0.15} & {4.97$\pm$0.23} &  {17.35} \\
{Multi-speaker FS2+\textit{d}-vector}  & {3.74$\pm$0.14} & {5.03$\pm$0.22} &  {17.55} \\

StyleSpeech & 3.77$\pm$0.15 & 5.01$\pm$0.23 & 17.51 \\

\bf Meta-StyleSpeech & \bf 3.82$\pm$0.15 & \bf 4.95$\pm$0.24 & \bf 16.79 \\
\bottomrule
\end{tabular}
\end{adjustbox}
\caption{MOS, MCD and WER for unseen speakers.}
\vspace{-0.1in}
\label{unseen_mos}
\end{table}

In the training process, we train StyleSpeech for 100k steps. For Meta-StyleSpeech, we start from pretrained StyleSpeech that is trained for 60k steps, and then meta-train the model for additional 40k steps via episodic training. We find that meta training from pretrained StyleSpeech helps obtain better stability in training. Furthermore, we use the predicted duration, pitch, and energy from the variance adaptor to generate the query speech in Meta-StyleSpeech training. In addition, we train our models with a minibatch size of 48 for StyleSpeech and 20 for Meta-StyleSpeech using the Adam optimizer. The parameters we use for the Adam optimizer are $\beta_1 = 0.9, \beta_2 = 0.98, \epsilon = 10^{-9}$. The learning rate of generator and mel-style encoder follows \citet{Vaswani2017AttentionIA}, while the learning rate of discriminator is fixed as 0.0002. We use MelGAN \citep{Kumar2019MelGANGA} as the vocoder to convert the generated mel-spectrograms into audio waveforms.

\begin{figure*}[t]
\centering
\begin{subfigure}{0.32\textwidth}
\centering
\includegraphics[width=0.9\linewidth]{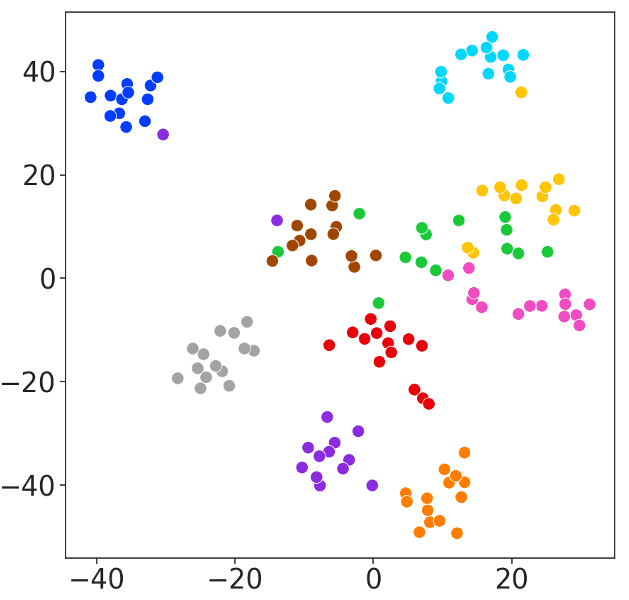}
\caption{Meta-StyleSpeech}
\end{subfigure}
\begin{subfigure}{0.32\textwidth}
\centering
\includegraphics[width=0.9\linewidth]{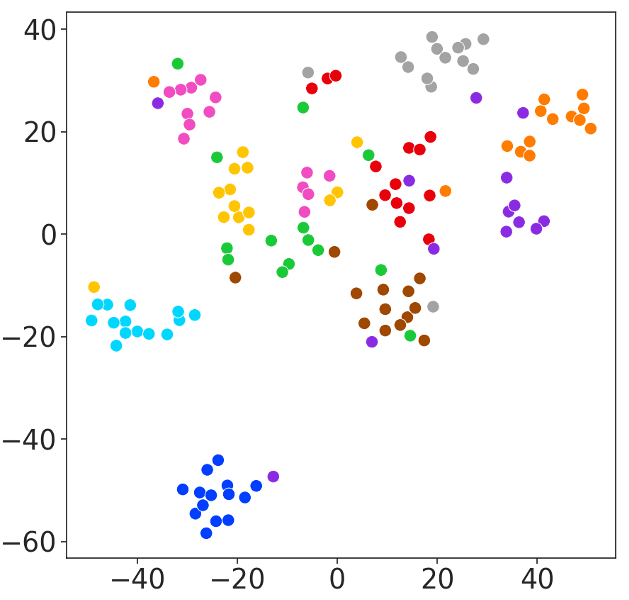}
\caption{StyleSpeech}
\end{subfigure}
\begin{subfigure}{0.32\textwidth}
\centering
\includegraphics[width=0.9\linewidth]{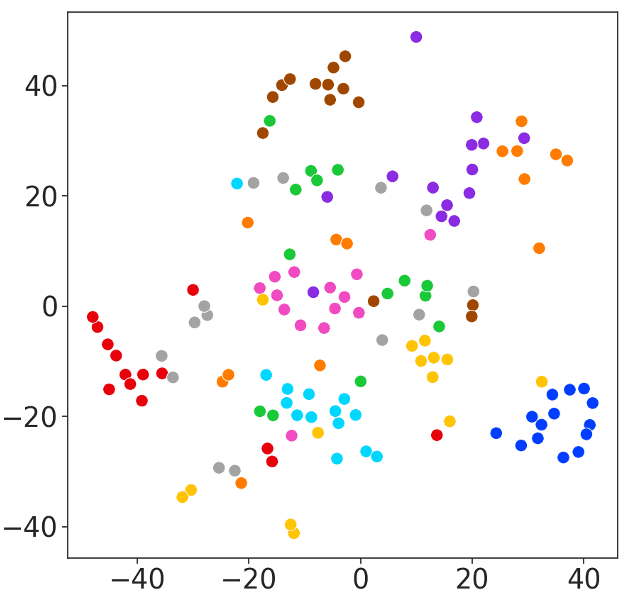}
\caption{GMVAE}
\end{subfigure}

\caption{TSNE visualization of the style vectors for unseen speakers (VCTK) with (a) and without (b) meta learning, and (c) GMVAE.}
\label{fig3}
\end{figure*}

\begin{table*}[t]
\centering
\resizebox{0.97\linewidth}{!}{
\renewcommand{\arraystretch}{0.75}
\renewcommand{\tabcolsep}{0.5mm}
\begin{tabular}{@{}ccccccccccccc@{}}
\toprule
\bf Metric          & \multicolumn{4}{c}{\textbf{SMOS ($\mathbf{\uparrow}$)}} & \multicolumn{4}{c}{\textbf{Sim ($\mathbf{\uparrow}$)}} & \multicolumn{4}{c}{\textbf{Accuracy ($\mathbf{\uparrow}$)}} \\ \cmidrule(l){2-13}
Length 
& \textless 1 sec & 1$\sim$3\,sec & 1 sen. & 2 sen.
& \textless 1 sec & 1$\sim$3\,sec & 1 sen. & 2 sen.
& \textless 1 sec & 1$\sim$3\,sec & 1 sen. & 2 sen. \\
\midrule
\midrule
{GMVAE}
& {2.85$\pm$0.12} &  {3.01$\pm$0.12} &  {2.91$\pm$0.16}  & {3.11$\pm$0.10} 
& {0.629} &  {0.695}  & {0.748}  & {0.765}
& {20.75\%}  & {30.49\%}  & {28.33\%} & {46.15\%}  \\
{Multi-speaker FS2(\textit{vanilla})}
& {3.14$\pm$0.17} &  {3.63$\pm$0.16} &  {3.31$\pm$0.14}  & {3.36$\pm$0.12} 
& {0.713} &  {0.735}  & {0.775}  & {0.773}
& {64.80\%}  & {73.80\%}  & {72.60\%} & {81.40\%}  \\
{Multi-speaker FS2+\textit{d}-vector}  
& {1.85$\pm$0.12}  &  {2.08$\pm$0.16} &  {2.11$\pm$0.16}  & {2.12$\pm$0.14} 
& {0.601} &  {0.603}  & {0.619}  & {0.616}
& {2.40\%}  & {3.80\%}  & {5.60\%} & {5.60\%}  \\
StyleSpeech    
& {3.32$\pm$0.16} &  {4.13$\pm$0.16} & \bf {3.50$\pm$0.10}  & {3.46$\pm$0.12}  
& 0.725 &  0.756  & 0.791  & 0.795
& 77.60\%  & 85.00\%  & 83.46\% & 85.19\%  \\
Meta-StyleSpeech  
& \bf {3.66$\pm$0.13} & \bf {4.19$\pm$0.14}  & {3.43$\pm$0.14}  & \bf {3.81$\pm$0.12}
& \bf 0.738   & \bf 0.779   & \bf 0.813  & \bf 0.815
& \bf 82.60\%   & \bf 90.20\%   & \bf 88.66\% & \bf 91.20\% \\           
\bottomrule
\end{tabular}
}
\caption{Adaptation performance on speech from unseen speakers with varying length of reference audios.} 
\label{unseen_smos}
\end{table*}

\begin{table}[t]
\centering
\small
\begin{tabular}{@{}ccccc@{}}
\toprule
\multicolumn{2}{c}{\bf Metric}           &  \textbf{MCD ($\mathbf{\downarrow}$)} & \textbf{Sim ($\mathbf{\uparrow}$)} & \textbf{Accuracy ($\mathbf{\uparrow}$)} \\ 
\midrule
\multirow{2}{*}{Gender} & Male       & 4.69$\pm$0.23   &  0.76   &  91\%      \\
                        & Female     & 4.87$\pm$0.23    & 0.75    & 89\%         \\ \midrule
\multirow{5}{*}{Accent} & American   &  4.80$\pm$0.23   & 0.77    &  91\%        \\
                        & British    &  4.83$\pm$0.21   & 0.75    &  91\%        \\
                        & Indian     &  5.28$\pm$0.31   & 0.74    &  86\%      \\
                        & African    &  5.06$\pm$0.24   & 0.76    &  93\%       \\
                        & Australian &  5.01$\pm$0.30   & 0.75    &  95\%       \\ \bottomrule
\end{tabular}
\caption{Adaptation performance depends on gender and accents.}
\label{genderandaccent}
\end{table}

\paragraph{Baselines} We compare our model with several baselines. \textbf{1) GT (\textit{oracle})}: This is the Ground-Truth speech. \textbf{2) GT mel (\textit{oracle})}: This is the speech synthesized by MelGAN vocoder using Ground-Truth mel-spectrogram. 3) \textbf{Deepvoice3}  \citep{Ping2017DeepV3}: This is a multi speaker TTS model which learns a look-up table to map embeddings for different speaker identity. Since DeepVoice3 is able to generate only seen speakers, we only compare against DeepVoice3 on seen speaker evaluation task. \textbf{4) GMVAE} \citep{Hsu2019HierarchicalGM}: This is a multi-speaker TTS model based on the Tacotron with a variational approach using Gaussian Mixture prior. \textbf{5) Multi-speaker FS2 (\textit{vanilla})}: This is a multi-speaker FastSpeech2, which adds the style vector to the encoder output and the decoder input. The style vector is extracted by the mel-style encoder. \textbf{6) Multi-speaker FS2 + \textit{d}-vector}: This is same as \textbf{5) Multi-speaker FS2 (\textit{vanilla})} except that the style vector is extracted from a pre-trained speaker verification model as suggested in \citet{Jia2018TransferLF}. \textbf{6) StyleSpeech:} This is our proposed model which generate multi-speaker speech from a single speech audio with the style-adaptive layer normalization and the mel-style encoder. \textbf{7) Meta-StyleSpeech:} This is our proposed model which extends StyleSpeech to perform meta training, with two additional discriminators to guide the generator. 

\paragraph{Evaluation Metric} The evaluation of the TTS models is very challenging, due to its subjective nature in the evaluation of the perceptual quality of generated speech. However, we use a sufficient number of metrics to evaluate the performance of our model. Specifically, for subjective evaluation, we conduct human evaluations with \textbf{MOS} (mean opinion score) for naturalness and \textbf{SMOS} (similarity MOS) for similarity. Both metrics are rated in 1-to-5 scale and reported with the 95\% confidence intervals (CI). 50 judges were participated in each experiment, where they were allowed to evaluate each audio sample once, that are presented in random orders.

In addition to subjective evaluation, we also conduct objective evaluation with quantitative measures. \textbf{MCD} evaluates the compatibility between the spectra of two audio sequences. Since the sequences are not aligned, we perform Dynamic Time Warping to align the sequences prior to comparison. In addition, \textbf{WER} validates an intelligibility of generated speech. We use a pre-trained ASR model DeepSpeech2 \citep{Amodei2016DeepS2} to compute Word Error Rate (WER). Note that both MCD and WER are not absolute metrics for evaluating speech quality, so we only use them for relative comparisons.

Beyond the quality evaluation, we also evaluate how similar the generated speech is, to the voice of the target speaker. In particular, we use the speaker verification model based on \citet{Wan2018GeneralizedEL} to extract x-vectors from generated speech as well as the actual speech of the target speaker. We then compute cosine similarity between them and denote it as \textbf{Sim}. The Sim scores are between -1 and 1, with higher scores indicating higher similarity. Since the verification model can be used for any speakers, we use it to evaluate both the trained speakers and new speakers.

Furthermore, we use a speaker classifier that identifies which speaker an audio sample belongs to. We train the classifier on VCTK dataset and the classifier achieve 99\% accuracy for validation set. We then calculate its \textbf{accuracy} to conduct the evaluation for adaptability of our model to new speakers. Better adaptation would result in higher classification accuracy.

\subsection{Evaluation on Trained Speakers}

Before investigating the ability of our model on unseen speaker adaptation, we first evaluate the quality of speech synthesized for seen (trained) speakers. To this end, we randomly draw one audio sample as reference speech for each 100 seen speakers in validation set of LibriTTS dataset. Then, we synthesize speech using the reference speech audio and the given text. Table \ref{seen_mos} shows the results of MOS, MCD, and WER evaluation on different models for the LibriTTS dataset. Our StyleSpeech and Meta-StyleSpeech outperform the baseline method in all three metrics, indicating that the proposed models synthesize higher quality speech. 

We also evaluate the similarity between synthesized speech for seen speakers and the reference speech. Table \ref{seen_smos} shows the results of SMOS and Sim score on different models for the LibirTTS dataset. Our StyleSpeech variants also achieve higher similarity scores to the reference speech than the other text-to-speech baseline methods. We thus conclude that our models are more effective in style transfer from reference speech samples. Furthermore, on both experiments, Meta-StyleSpeech achieves the best performance, which shows the effectiveness of the meta-learning. 

\subsection{Unseen Speaker Adaptation}

\begin{figure}[t]
\centering
\begin{subfigure}{0.835\linewidth}
\centering
\includegraphics[width=\linewidth]{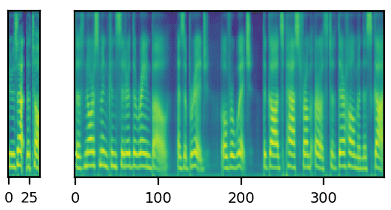}
\caption{\textbf{Left:} A reference speech sample, \textbf{Right:} Generated mel-spectrogram}
\end{subfigure}
\begin{subfigure}{0.68\linewidth}
\centering
\includegraphics[width=\linewidth]{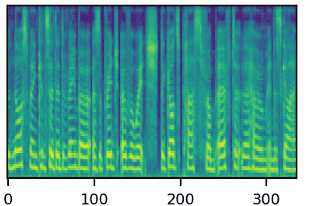}
\caption{Ground truth mel-spectrogram}
\end{subfigure}
\caption{Mel-spectrogram generation with a reference speech shorter than 1 second.}
\label{fig4}
\vspace{-0.1in}
\end{figure}

We now validate the adaptation performance of our model on unseen speakers. To this end, we first evaluate the quality of generated speech of unseen speakers. In this experiment, we also randomly draw one audio sample as reference for each 108 unseen speaker in the VCTK datasets and then generate speech using the reference speech and the given text. Table \ref{unseen_mos} shows the results of MOS, MCD and WER evaluation on speech from unseen speakers. Meta-StyleSpeech also achieves the best generation quality in all three metrics, largely outperforming the baselines. 

As done in the seen-speaker experiments, we also evaluate the similarity to reference speech for unseen speakers. Following \citet{Nachmani2018FittingNS}, we expect that the ability to adapt to new speakers may depend on the length of the reference speech audio. Thus, we perform the experiments while varying the length of the reference speech audio from unseen speakers. We use four different lengths: \textless 1 sec, 1$\sim$3 sec, 1 sentence and 2 sentences. We set 1 sentence as a speech audio sample which is longer than 3 seconds in length and, for 2sentences, we simply concatenate two speech audio samples. 

The results of SMOS, Sim, and Accuracy are presented in Table \ref{unseen_smos}. As shown in the result, our model, Meta-StyleSpeech,  significantly outperforms the baseline as well as StyleSpeech on reference speech of any lengths. Specifically, it achieves high adaptation performance even with the reference speech that is shorter than 1 second. Figure \ref{fig4} shows the example of generated speech from the reference speech shorter than 1 second, using Meta-StyleSpeech. We observe that our model generates high quality speech with sharp harmonics and well resolved formants which is comparable to ground-truth mel-spectrogram.

In addition, we also conduct adaptation evaluation depends on various attributes such as gender and accent. The result is shown in Table \ref{genderandaccent}. We can see that Meta-StyleSpeech shows balanced results for gender. Moreover, the model also shows high adaptation performance for all accents. However, we can see that slightly lower performance on Indian accent. This could be because the Indian accent have more dramatic variation in speaking style than other accents.  

\begin{table}[t]
\small
\centering
\begin{adjustbox}{width=0.98\linewidth}
\begin{tabular}{@{}cccc@{}}
\toprule
            & \begin{tabular}[c]{@{}c@{}}\bf MCD ($\mathbf{\downarrow}$)\\ (seen)\end{tabular} & \begin{tabular}[c]{@{}c@{}}\bf MCD ($\mathbf{\downarrow}$)\\ (unseen)\end{tabular} & \begin{tabular}[c]{@{}c@{}}\bf Accuracy ($\mathbf{\uparrow}$)\\ (\textless{}1sec)\end{tabular} \\ 
\midrule
\midrule
Meta-StyleSpeech
    & \bf 4.29$\pm$0.21  
    & 4.95$\pm$0.24  
    & \bf 82.60\% \\
StyleSpeech 
    & 4.49$\pm$0.22  
    & 5.01$\pm$0.23  
    & 77.60\% \\
w/o $D_t$   
    & 4.85$\pm$0.21                         
    & 5.53$\pm$.27               
    & 78.00\%     \\
w/o $D_s$   
    & 4.51$\pm$0.20                       
    & 5.17$\pm$0.24                      
    & 80.40\%     \\
w/o $\mathcal{L}_{cls}$   
    & 4.32$\pm$0.20                        
    & \bf 4.85$\pm$0.24                      
    & 80.60\%     \\

\bottomrule
\end{tabular}
\end{adjustbox}
\caption{Ablation study for verifying the effectiveness of the phoneme discriminator, the style discriminator and the style prototypes.}
\label{ablation}
\vspace{-0.18in}
\end{table}

\paragraph{Visualization of style vector} To better understand effectiveness of meta-learning, we visualize the style vectors. In figure \ref{fig3}, we demonstrate the t-SNE projection \citep{Maaten2008VisualizingDU} of style vectors from unseen speakers from the VCTK datasets. In particular, we select 10 female speakers who have similar accents and voices which are difficult to  distinguish by the model. We can see that while StyleSpeech clearly better separates the style vectors when compared with GMVAE, Meta-StyleSpeech trained with meta-learning achieves even better clustered style vectors.

\subsection{Ablation Study}
We further conduct an ablation study to verify the effectiveness of each components in our model, including the text discriminator, the style discriminator and the style prototypes. In the ablation study, we use two metrics; MCD and Accuracy. The results are shown in Table \ref{ablation}. Both the quality of generated speech and adaptation ability of the model are significantly dropped when removing the text discriminator. This indicate that the role of the text discriminator is very important in meta-training. We also find that removing the style discriminator and the style prototypes result in the performance drop on unseen speaker adaptation. From this result, we find that the style discriminator is important in helping the model adapt to speech from unseen speakers and the style prototypes further enhance the adaptation performance of the model.

\section{Conclusion}
We have proposed StyleSpeech, a multi-speaker adaptive TTS model which can generate high quality and expressive speech from a single short-duration audio sample of the target speaker. In particular, we propose a style-adaptive layer normalization (SALN) to generate various styles of speech of multiple speakers. Furthermore, we extend StyleSpeech to Meta-StyleSpeech with additional discriminators and meta-learning, to improve its adaptation performance on unseen speakers. Specifically, we simulate one-shot episodic training and the style discriminator utilizes a set of style prototypes to enforce the generator to generate speech that follows the common style of each speaker. The experimental results demonstrate that StyleSpeech and Meta-StyleSpeech can synthesize high-quality speech given the reference audios from both seen and unseen speakers. Moreover, Meta-StyleSpeech achieves significantly improved adaptation performance on the speech from unseen speakers, even with a single reference speech with the length of less than one second. For future work, we plan to further improve Meta-StyleSpeech to perform controllable speech generation by disentangling its latent space, to enhance its practicality in diverse real-world applications.

\paragraph{Acknowledgements} This work was supported by Institute of Information \& communications Technology Planning \& Evaluation (IITP) grant funded by the Korea government (MSIT) (No.2019-0-00075, Artificial Intelligence Graduate School Program (KAIST)), and the Engineering Research Center Program through the National Research Foundation of Korea (NRF) funded by the Korean Government MSIT (NRF-2018R1A5A1059921). We sincerely thank the anonymous reviewers for their constructive comments which helped us significantly improve our paper during the rebuttal period.


\bibliography{References}
\bibliographystyle{icml2021}

\appendix

\clearpage
\onecolumn
\begin{center}{\bf {\LARGE Supplementary Material} }
\end{center}
\begin{center}{\bf {\Large Meta-StyleSpeech : Multi-Speaker Adaptive Text-to-Speech Generation} \linebreak}
\end{center}

\section{Detailed model architectures}

\begin{figure*}[h]
    \centering
    \advance\leftskip1.1cm
    \includegraphics[width=0.71\textwidth]{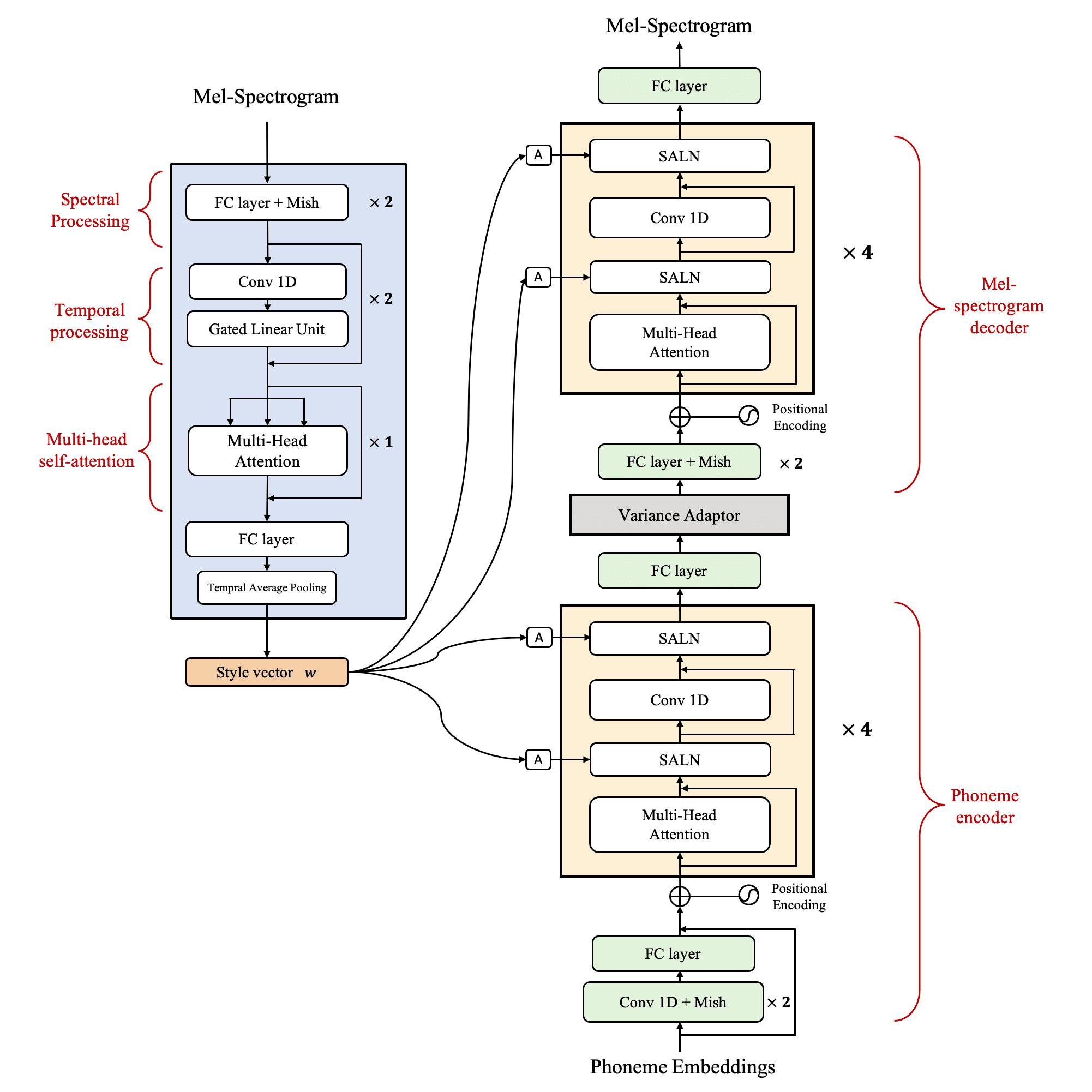}
    \caption{\textbf{Mel-Style Encoder and Generator. Mel-Style Encoder(Left): } The mel-style encoder is comprised of the spectral processing, temporal processing and multi-head self-attention modules. Spectral processing module consists of two fully-connected layers each of which has 128 hidden units. Temporal processing consists of two gated 1D CNNs with a residual connection. The filter size and the kernel size of 1D convolution layers are 128 and 5, respectively. Furthermore, the hidden size and number of attention heads in multi-head self-attention are 128 and 2. On top of the multi-head self-attention module, the final fully-connected layer has a dimensionality of 128, which has the same size with the style vector, followed by temporal average pooling. \textbf{Generator(Right): } A sequence of 256-D phoneme embeddings are fed into two 1D convolution layers followed by a fully-connected layer with residual connection. The filter size and the kernel size of 1D convolution layers are 256 and 3, respectively. Next, after positional encoding, the output go through 4 FFT blocks, a fully-connected layer and the variance adaptor. After the variance adaptor come two fully-connected layers which have [128, 256] units respectively, followed by positional encoding. The output then goes through 4 FFT blocks and the last fully-connected layer which have 80 units to match the dimension of mel-spectrogram. The hidden size, number of attention  heads, the kernel size and filter size of the 1D convolution in the FFT block are 256, 2, 9 and 1024, respectively. We use Mish activation in all layers of both the mel-style encoder and the generator. The dropout rate is 0.1.}
    \label{figure5}
\end{figure*}

\begin{figure*}[t]
    \centering
    \includegraphics[width=0.55\textwidth]{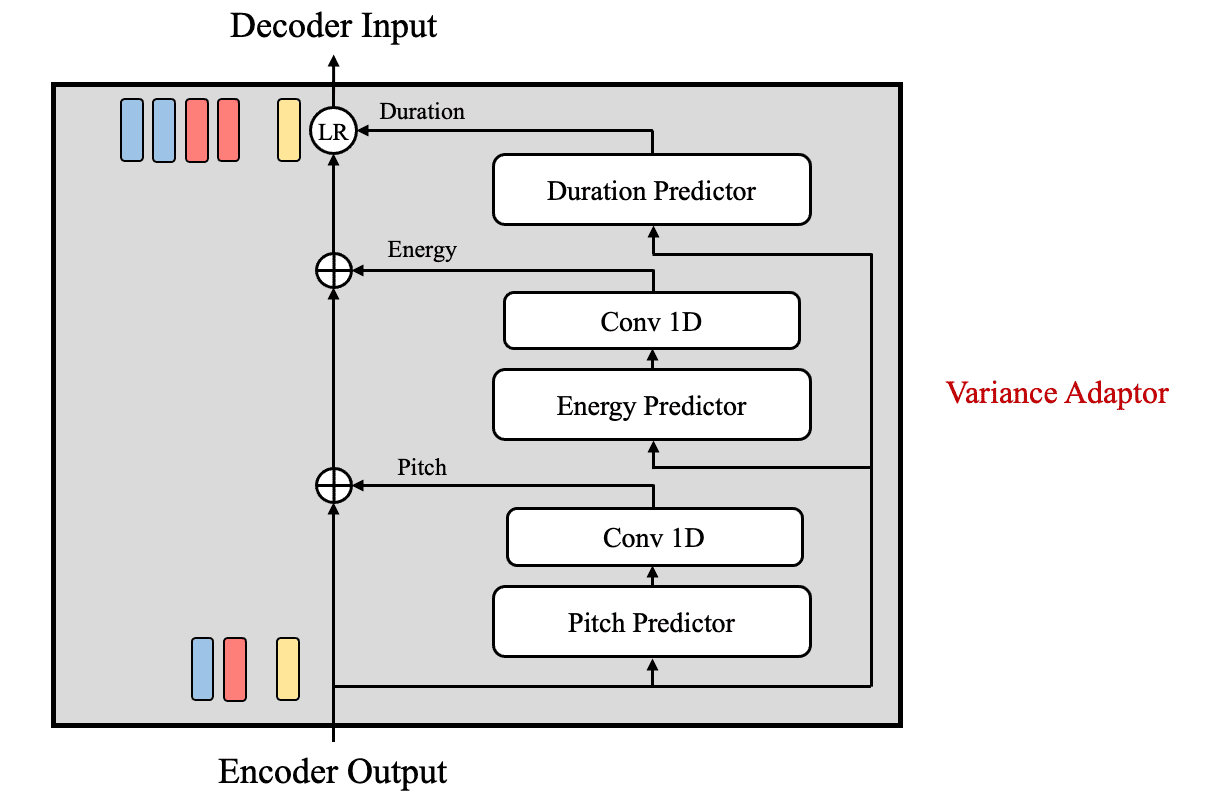}
    \caption{\textbf{Variance Adaptor.} The variance adaptor is comprised of a pitch predictor, an energy predictor and a duration predictor. The pitch, energy, and duration predictors have same architectures as FastSpeech2 which consist of 1D convolution, ReLU activation, and layer norm. However, we use a 1D convolution layer to add real or predicted pitch and energy directly to the encoder output. The kernel size of the 1D convolution is 9 and the filter size is 256 to match the hidden size of the encoder output.
    }
    \label{figure6}
\end{figure*}

\begin{figure*}[h!]
    \centering
    \includegraphics[width=0.75\textwidth]{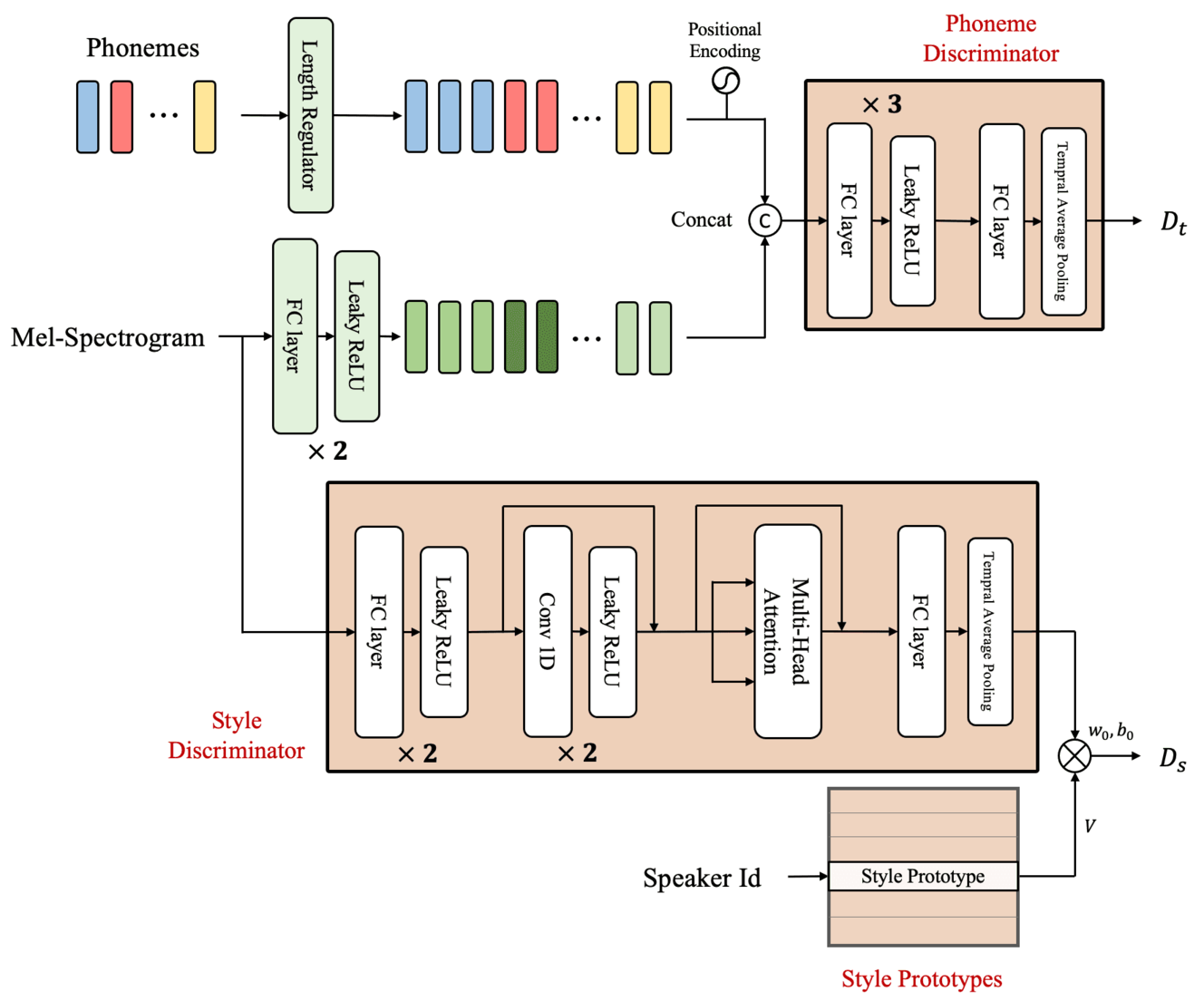}
    \caption{\textbf{Phoneme Discriminator and Style Discriminator.} \textbf{Phoneme Discriminator(Up): } The mel-spectrogram is fed into two fully-connected layers with 256 and concatenated with phoneme embeddings which pass through the length regulator followed by positional encoding. The concatenated output then passes through four fully-connected layers which have [512, 512, 512, 1] hidden units respectively, followed by temporal average pooling. \textbf{Style Discriminator(Down): } The style discriminator has a similar architecture with the mel-style encoder in Figure \ref{figure5}. The difference is that the style discriminator uses simple 1D convolutions instead of gated convolutions. The style prototypes have same size with the style vectors. Furthermore, we use Leaky-Relu for all the activation layers in both the phoneme discriminator and the style discriminator.}
    \label{figure7}
\end{figure*}

In this section, we describe the detailed architectures of our models. Specifically, Figure \ref{figure5} depicts both the mel-style encoder and generator, Figure \ref{figure6} depicts the variance adaptor and Figure \ref{figure7} depicts discriminators, the phoneme discriminator and the style discriminator.
\clearpage

\clearpage
\section{Meta-training Algorithm}
In this section, we provide the pseudo-code of our meta-learning algorithm to train Meta-StyleSpeech. Algorithm \ref{algorithm1} presents the episodic learning process (line 3-19). In each episode, we firstly select $N$ speakers from the entire set of speakers from the given training dataset, and sample one support sample $(X_s, t_s)$ and one query text $t_q$ for each speaker (line 4-7). We then update our models in two steps. Step \Romannum{1} updates the mel-style encoder and the generator (line 8-12). Step \Romannum{2} updates the phoneme discriminator, the style discriminator and the style prototypes (line 13-18).

\begin{figure*}[h!]
\begin{algorithm}[H]
\caption{Meta-training for StyleSpeech. $K$ is the number of speakers in the training set. $\theta$ denotes the parameters of the generator($G$). $\phi$ denotes the parameters of the mel-style encoder($Enc_s$). $\psi_1, \psi_2$ denote the parameters of the text discriminator($D_t$) and the style discriminator($D_s$), respectively.}
\label{algorithm1}
\begin{algorithmic}[1]
    \STATE Initialize parameters $\theta, \phi, \psi_1, \psi_2$
    \STATE Initialize style prototypes $S=\{s_i\}_{i=1}^K$, where $s_i$ denotes style prototype of the speaker $i$.
    
    \WHILE {not done} 
    \STATE Sample $N$ speakers from $\{1,\dots,K\}$
    \FORALL{$k\in\{1,\dots,N\}$}
    \STATE Sample support sample and query text $\{X_s, t_s, t_q\}$ correspond to $k$
    \ENDFOR
    
    \STATE Step \Romannum{1}. Update $Enc_s$ and $G$.
    \STATE $w_s \leftarrow Enc_s(X_s)$
    \STATE $\mathcal{L}_{adv} \leftarrow (D_s(G(t_q,w_s), s_k)-1)^2 + (D_t(G(t_q,w_s), t_q)-1)^2$
    \STATE $\mathcal{L}_{recon} \leftarrow \norm{G(t_s, w_s) - X_s}_1$
    \STATE Update $\theta,\phi$ using $\mathcal{L}_G(t_q, w_s, t_s, s_k) = 10\mathcal{L}_{recon} + \mathcal{L}_{adv}$.
    
    \STATE Step \Romannum{2}. Update $D_t$ and $D_s$.
    \STATE $\mathcal{L}_{D_t} \leftarrow (D_t(X_s, t_s)-1)^2 + D_t(G(t_q,w_s), t_q)^2$
    \STATE $\mathcal{L}_{cls} \leftarrow -\log \, \frac{\exp(w_s^Ts_k))}{\sum_{k^{\prime}} \exp(w_s^T s_{k^{\prime}})} $
    \STATE $\mathcal{L}_{D_s} \leftarrow (D_s(X_s, s_k)-1)^2 + D_s(G(t_q,w_s), s_k)^2$
    \STATE Update $\psi_1,\psi_2$ using $\mathcal{L}_D(t_q, w_s, s_k) = \mathcal{L}_{D_t} + \mathcal{L}_{D_s} + \mathcal{L}_{cls} $
  \ENDWHILE
\end{algorithmic}
\end{algorithm}
\end{figure*}

\clearpage
\section{MOS evaluation interface} 
\begin{figure*}[h!]
    \centering
    \advance\leftskip0.8cm
    \includegraphics[width=0.95\textwidth]{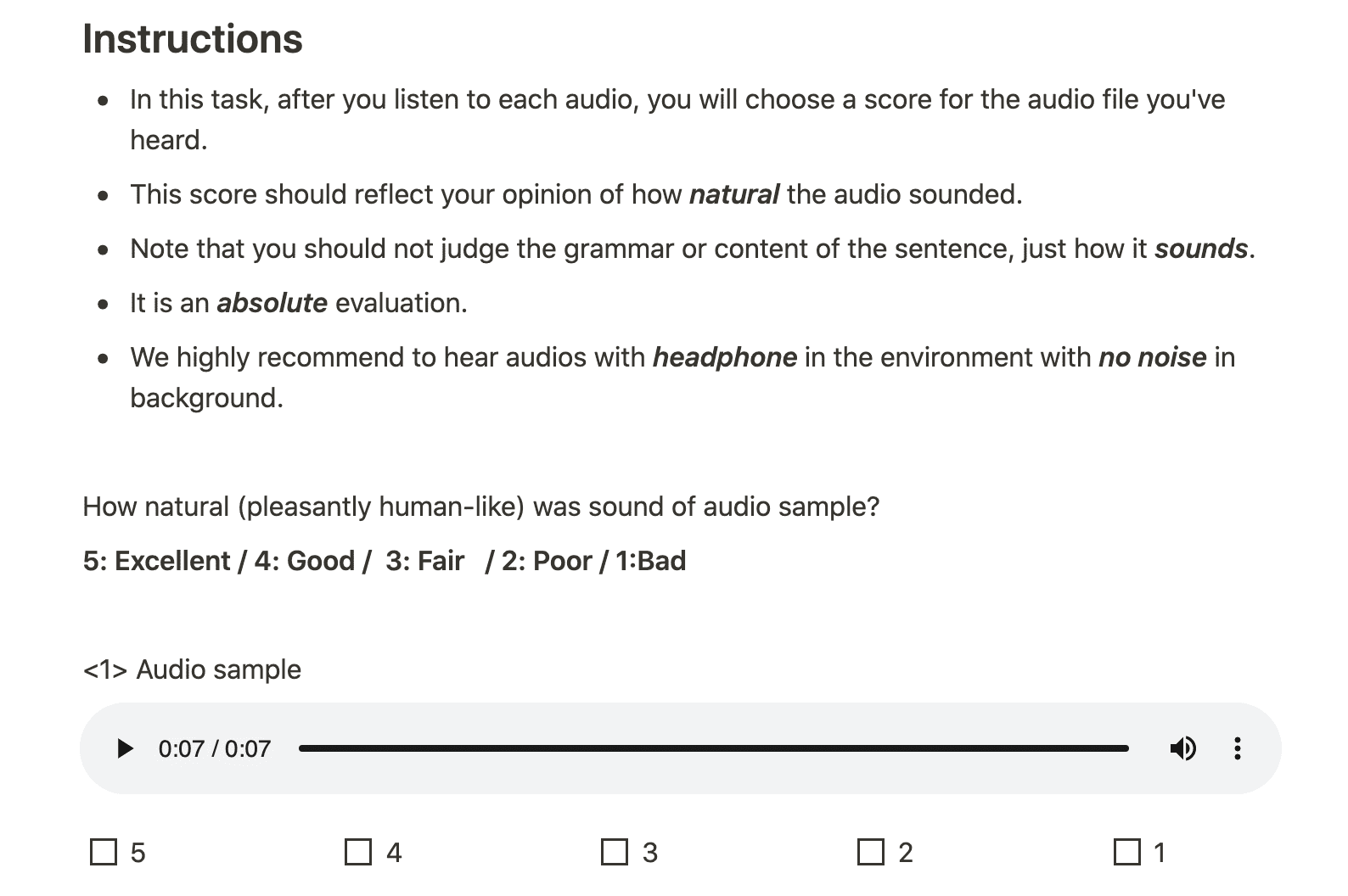}
    \caption{\textbf{Interface of MOS evaluation for naturalness.}}
    \label{figure8}
\end{figure*}




\end{document}